\documentclass[preprintnumbers,prd,superscriptaddress,nofootinbib,floatfix,showpacs,showkeys]{revtex4}

\usepackage{amsmath}
\usepackage{amsfonts}
\usepackage{amssymb}
\usepackage{graphicx}
\usepackage{color}
\usepackage{hyperref}
\usepackage{booktabs}
\usepackage{cleveref}
\usepackage{braket}
\usepackage{mathtools}
\usepackage[rightcaption]{sidecap}
\usepackage{subfigure}
\usepackage{comment}

\definecolor{vividviolet}{rgb}{0.62, 0.0, 1.0}
\definecolor{amaranth}{rgb}{0.9, 0.17, 0.31}
\definecolor{palatinateblue}{rgb}{0.15, 0.23, 0.89}
\definecolor{brightpink}{rgb}{1.0, 0.0, 0.5}
\definecolor{cornflowerblue}{rgb}{0.39, 0.58, 0.93}
\definecolor{deepcarminepink}{rgb}{0.94, 0.19, 0.22}
\definecolor{radicalred}{rgb}{1.0, 0.21, 0.37}
\hypersetup{ linktoc=all,
    colorlinks, linkcolor={palatinateblue},
    citecolor={brightpink}, urlcolor={amaranth}
}

\newcommand{\be}{\begin{equation}}
\newcommand{\ee}{\end{equation}}
\newcommand{\bs}{\begin{split}}
\newcommand{\bea}{\begin{eqnarray}}
\newcommand{\eea}{\end{eqnarray}}

\newcommand{\bes}{\begin{subequations}}
\newcommand{\ees}{\end{subequations}}

\graphicspath{{Graphics/}}

\def\be{\begin{equation}}
\def\ee{\end{equation}}
\def\bea{\begin{eqnarray}}
\def\eea{\end{eqnarray}}

\def\nat{Nature}

\def\prd{Phys. Rev. D}
\def\mnras{MNRAS}
\def\aj{AJ}
\def\apj{ApJ}

\def\apjs{ApJ Suppl. Ser.}

\def\physrep{Phys. Rep.}
\def\jcap{JCAP}

\def\apss{Astrophysics and Space Science}

\begin{document}

\title{Double polytropic cosmic acceleration from the Murnaghan equation of state}

\author{Peter K. S. Dunsby}
\email{peter.dunsby@uct.ac.za}
\affiliation{Department of Mathematics and Applied Mathematics, University of Cape Town, Rondebosch 7701, Cape Town, South Africa.}
\affiliation{Cosmology and Gravity Group (CGG),
University of Cape Town, Rondebosch 7701, Cape Town, South Africa.}
\affiliation{South African Astronomical Observatory, Observatory 7925, Cape Town, South Africa.}
\affiliation{Centre for Space Research, North-West University, Potchefstroom 2520, South Africa}

\author{Orlando~Luongo}
\email{orlando.luongo@unicam.it}
\affiliation{Universit\`a di Camerino, Via Madonna delle Carceri 9, 62032 Camerino, Italy.}
\affiliation{SUNY Polytechnic Institute, 13502 Utica, New York, USA.}
\affiliation{Istituto Nazionale di Fisica Nucleare, Sezione di Perugia, 06123, Perugia,  Italy.}
\affiliation{INAF - Osservatorio Astronomico di Brera, Milano, Italy.}
\affiliation{Al-Farabi Kazakh National University, Al-Farabi av. 71, 050040 Almaty, Kazakhstan.}

\author{Marco Muccino}
\email{marco.muccino@lnf.infn.it}
\affiliation{Universit\`a di Camerino, Via Madonna delle Carceri 9, 62032 Camerino, Italy.}
\affiliation{Al-Farabi Kazakh National University, Al-Farabi av. 71, 050040 Almaty, Kazakhstan.}
\affiliation{ICRANet, P.zza della Repubblica 10, 65122 Pescara, Italy.}

\author{Vineshree Pillay}
\email{vineshree.pillay013@gmail.com}
\affiliation{Department of Mathematics and Applied Mathematics, University of Cape Town, Rondebosch 7701, Cape Town, South Africa.}
\affiliation{Cosmology and Gravity Group (CGG),
University of Cape Town, Rondebosch 7701, Cape Town, South Africa.}

\begin{abstract}
We consider a double polytropic cosmological fluid and  demonstrate that, when one constituent resembles a bare cosmological constant while the other emulates a generalized Chaplygin gas, a good description of the Universe's large-scale dynamics is obtained. In particular, our double polytropic reduces to  the Murnaghan equation of state, whose applications are already well established in solid state physics and classical thermodynamics. Intriguingly, our model approximates the conventional $\Lambda$CDM paradigm while reproducing the collective effects of logotropic and generalized Chaplygin fluids across different regimes. To check the goodness of our fluid description, we analyze first order density perturbations, refining our model through various orders of approximation, utilizing  $\sigma_8$ data alongside other cosmological data sets. Encouraging results suggest that our model, based on the Murnaghan equation of state, outperforms the standard cosmological background within specific approximate regimes and, on the whole, surpasses the standard phenomenological reconstruction of dark energy.
\end{abstract}



\maketitle

\section{Introduction}
The current cosmic speed-up, supported by several observations \cite{1998Natur.391...51P,1998AJ....116.1009R,1999ApJ...517..565P,2003ApJ...594....1T,2003Sci...299.1532B,2003ApJS..148....1B,2003ApJS..148..135H,2003ApJS..148..161K,2003ApJS..148..175S,2005ApJ...633..560E}, is widely attributed to an unknown exotic fluid with a negative equation of state (EoS) \cite{2000IJMPD...9..373S,2006IJMPD..15.1753C,tsujikawa2011dark,2006IJMPD..15.1753C,2009IJMPA..24.1545M,2012Ap&SS.342..155B}, well described by the cosmological constant $\Lambda$ purported by the concordance model \cite{2003PhR...380..235P} or, more broadly, by some sort of dark energy contribution \cite{2003RvMP...75..559P}, behaving anti-gravitationally \cite{Luongo:2014qoa,Giambo:2020jjo,Luongo:2023xaw,Luongo:2023aib}.
The dark energy nature is still poorly understood \cite{capozziello2013cosmographic}, but several models have been proposed either to explain the cosmic acceleration    \cite{2009PhRvD..80b3008L,2020EPJC...80..629P,2003PhRvD..67f3504B,2007IJGMM..04..209S,2017PhR...692....1N,2018FoPh...48...17L,2013JCAP...07..017B}, or to solve the conceptual and theoretical issues raised by observed cosmological tensions \cite{weinberg1989cosmological,2018PhRvD..98j3520L,gruber2014cosmographic}.

Dark matter is a completely different component  that represents a further challenge for modern cosmology \cite{del2012three}, specifically the identification of its particle candidates \cite{2021PrPNP.11903865A}, among which we include ultralight fields, extremely massive particles, geometric contributions, or proposed extensions of Einstein's gravity \cite{2021PrPNP.11903865A,2019Univ....5..213P,Capozziello:2018hly}, without conclusive evidences in favour of any of them\footnote{For alternative interpretations toward dark matter, refer to Ref. \cite{Belfiglio:2024xqt,Luongo:2023aaq,Belfiglio:2023moe,Belfiglio:2023rxb,Belfiglio:2022yvs,Belfiglio:2022cnd}.} \cite{2020Symm...12.1648P}.

Recently, a number of unified theories \cite{avelino2008linear} have been proposed to unify both dark energy and dark matter into a single dark fluid \cite{becca2007dynamics} with a negative net pressure able to drive the cosmic acceleration \cite{bento1999compactification} and treat the dark sector in the same way at the perturbative level for structure formation \cite{hu1999structure}.
However, unified dark energy models like dark fluids \cite{2011PhRvD..84h9905A} and logotropic models \cite{2015EPJP..130..130C} are still incomplete and have been severely criticised \cite{2011PhLB..694..289F,2014IJMPD..2350012L,2018PhRvD..98j3520L,2021PhRvD.104b3520B}.

In order to overcome the above issues and construct a unified dark energy model, a double field approach in which a tachyonic field is minimally-coupled to a further scalar, carrying vacuum energy, has been proposed in Ref. \cite{Dunsby:2023qpb}. This scenario leads to a single effective fluid that behaves pressure-less at early times, while providing a negative pressure at late times. The advantage of this recipe is that the tachyonic field behaves similarly to previous unified dark energy models, like the Chaplygin gas model \cite{2002PhRvD..66d3507B}, but also includes quantum fluctuations due to a symmetry breaking mechanism associated with the field transporting vacuum energy \cite{Dunsby:2023qpb}.

In this paper, we notice that a possible macroscopic interpretation of the above unified scalar fields has been phenomenologically obtained within the context of solid state physics through the introduction of the {\it Murnaghan EoS} \cite{murnaghan1944compressibility}. It follows that our effective  cosmological matter is no longer a dust-like fluid, thus  affecting the clustering of structures in a different way than the $\Lambda$CDM model during early times. Hence, physical consequences of the fact that the Murnaghan EoS yields the concept of matter with pressure are thus investigated. To do so, we propose a double polytropic interpretation of the Murnaghan EoS and show that it can reduce to the $\Lambda$CDM, logotropic and generalized Chaplygin gas (GCG) models at different epochs of the Universe evolution. Particularly, we first constrain the model free parameters and assess its compatibility with observations. Accordingly, we work with Markov chain Monte Carlo (MCMC) simulations, based on the Metropolis-Hastings algorithm and employ the most recent low-redshift $z$ catalogs such as the observational Hubble data (OHD) \citep{2022LRR....25....6M}, type Ia supernovae (SNe Ia) \cite{2018ApJ...859..101S}, the observed growth function and the RMS mass fluctuations \cite{Paul}. Afterwards, we compare our findings with the standard $\Lambda$CDM model. Then, using our outcomes, we study the thermodynamic behaviour of the Murnaghan fluid to compare and contrast it with other unified dark energy models. Since the Murnaghan EoS exhibits a ``chameleon" behavior at different scales, we show its adaptability and propose the concept of matter with pressure as a robust candidate to describe the large-scale dynamics.

The paper is structured as follows. In Section \ref{sezione2} we introduce the Murnaghan EoS and its limiting cases. Section \ref{sezione3} deals with the numerical analysis and Section \ref{sezione4} with the physical interpretation of the results. Section \ref{sezione6} shows the impact of matter with pressure in linear perturbations. Finally in Section \ref{sezione7} we conclude this work and summarize our conclusions.

\section{The Murnaghan equation of state} \label{sezione2}
In studies of stellar structure, polytropic EOSs are widely used because they suggest specific forms of heat capacities for astrophysical compact objects.
In recent years, polytropic fluids have also been applied within the context of cosmological modeling with promising results \cite{Muccino:2020gqt}. Specifically, the impact on structure formation can be investigated by assuming a single-fluid polytropic EoS of the form,
\begin{equation}\label{polyeos}
P=P_0\rho^{1+\gamma}\,,
\end{equation}
where $\gamma=1/n$, $n\in \mathbb{N}$, the polytropic index, and, by virtue of the adiabatic sound speed defined as,
\begin{equation}\label{suono}
c_s^2\equiv\frac{\partial P}{\partial \rho},
\end{equation}
one immediately obtains through Eq. \eqref{polyeos},
\begin{equation}\label{sonido}
c_s^2=P_0(1+\gamma)\rho^\gamma\neq0\,.
\end{equation}
This clearly cannot be zero since $\gamma\neq-1$ and, so, this fact clearly fixes a non-zero Jeans length, reducing the regions to form structure.

However, a genuine cosmological constant, defined as $\gamma\rightarrow-1$, would imply $c_s=0$, leading to an always zero Jeans length. We also notice that when $\gamma=-2$ the EoS resembles the Chaplygin gas model $P=P_0/\rho$, further reinforcing the idea that relaxing the hypothesis $\gamma=1/n$ would open new avenues in the physics of polytropes.

Motivated by the above considerations, a different perspective can arise if one considers a double polytropic fluid where one can introduce an additional parameter to account for the behavior of dark energy at different energy densities. This is identified by the following relation \cite{2016PhRvD..94h3525D},
\begin{equation}
P=P_1\rho^{1+\gamma_1}+P_2\rho^{1+\gamma_2}\,,
\end{equation}
with $P_1,P_2,\gamma_1,\gamma_2$ denoting unbounded constants related to different behaviours of the fluid at different energy density regimes, whose signs are unfixed.
Adopting Eq. \eqref{suono}, we now obtain,
\begin{equation}
c_s^2=P_1(1+\gamma_1)\rho^{\gamma_1}+P_2(1+\gamma_2)\rho^{\gamma_2},
\end{equation}
and notice that there exists a region where, for certain values of the free parameters, the squared of the sound speed equals zero, in analogy to dust.

This model falls into the set of \emph{unified dark energy models}, made up by one single fluid only that behaves differently throughout the Universe evolution, mimicking both dark matter and dark energy at the same time. Through this interpretation, we find that the corresponding EoS is zero for stages of the Universe's evolution and deviates in this manner from the standard dust-like assumption for matter.

We can easily construct the simplest case found in solid state physics, known as the \emph{Murnaghan fluid}, by setting $\gamma_1<0$, but $\neq-1$, and $\gamma_2=0$, while identifying $P_2$ with the constant value of some sort of vacuum energy\footnote{For the sake of clearness, the constant is related to the bare cosmological constant. Its value might, in fact, agree with current observations and thus cannot be associated with primordial quantum fluctuations of scalar fields, expected to lie around  Planck energy scales. More precisely, the vacuum energy is associated with primordial quantum fluctuations and intimately related to the cosmological constant problem --- specifically, the inability to cancel out the degrees of freedom related to high-energy scales associated with quantum fluctuations before and after a primordial phase transition \cite{2012CRPhy..13..566M}.}, to have \cite{Dunsby:2023qpb},
\begin{equation}\label{nuovaeos}
P=\frac{P_1}{\rho^\beta}+P_2\,,
\end{equation}
where $\beta\equiv |\gamma_1|$. In other words, assuming Eq. \eqref{nuovaeos} implies the existence of a matter-like fluid that for very large densities behaves as a constant.
We find that the double polytropic EoS is analogous to the EoS given for a Murnaghan fluid, which describes the incompressibility of matter under high pressure  \cite{1944PNAS...30..244M}.

Hence, we assume that matter in the universe provides a bulk modulus of incompressibility at constant temperature $T$, given by $K=-V(\partial P/\partial V)_T=K_0+K_0^\prime P$, with $K^\prime\equiv\partial K/\partial P$, that establishes a relationship between the volume $V$ and the pressure $P$ of the universe through
\begin{equation}\label{P}
    P = \frac{K_0}{K_0^\prime}\left[\left(\frac{V}{V_0}\right)^{-K_0^\prime}-1\right]\,,
\end{equation}
where the subscript ``$0$'' denotes when $P=0$.
To well adapt this recipe to the dynamics of the Universe, we additionally require that
\begin{itemize}
\item[-] $P=0$ occurs when pressure-less matter dominates, namely, at a density $\rho_\star$ or at volume $V_0\propto\rho_\star^{-1}$;
\item[-] as $P\neq0$, the volume $V\propto\rho^{-1}$ is larger than $V_0$, or $\rho_\star>\rho$, to ensure late time expansion;
\item[-] $K>0$, thus $K_0>0$, when $P=0$, and $K_0^\prime<0$, when $P<0$ to ensure the accelerated expansion of the Universe.
\end{itemize}
We can change the variables from the original EoS such that $K_0\rightarrow A_\star$ and $K_0^\prime\rightarrow -\alpha$, with $\alpha$ and $A_\star$ being positive constants, to obtain a GCG, $P\propto \rho^{-\alpha}$,  with an extra constant \cite{2002PhRvD..66d3507B}, resulting in the recast version of Eq. \eqref{nuovaeos} given by,
\begin{equation}\label{P1}
    P=-\frac{A_\star}{\alpha}\left[\left(\frac{\rho_\star}{\rho}\right)^{\alpha}-1\right]\,.
\end{equation}
The dark fluid density in terms of the scale factor, $a=(1+z)^{-1}$, can be obtained  by solving the continuity equation
\begin{equation}
\frac{d\rho}{da} + \frac{3}{a}(P+\rho)=0\,.
\end{equation}
However, since the solution is not analytical, thus, to constrain $A_\star$, $\alpha$, and $\rho_\star$, we then explore below  some approximate regimes, directly coming from Eq. \eqref{P1}.
\begin{itemize}
\item[-]  {\bf The $\Lambda$CDM approximation:}\\ At redshift $z\approx0$, or $a= a_0\approx1$, $\rho$ approaches the critical value $\rho_c=3H_0^2/(8\pi G)$, where $H_0$ is the Hubble constant and $G$ is the gravitational constant.
Thus, for $\alpha\gtrsim0$, we obtain,
\begin{subequations}
\label{apprLCDM}
\begin{align}
\label{appr1}
\rho(a)&\approx \rho_c \left(\Omega_{\rm DF}+ \frac{P}{\rho_c}\right) a^{-3} - P\,,\\
\label{appr2}
P&\approx -A_\star \ln\left(\frac{\rho_\star}{\rho_c}\right)<0\,,
\end{align}
\end{subequations}
where $\Omega_{\rm DF}\equiv\rho/\rho_c$.
Including pressure-less baryons $\Omega_{\rm b}a^{-3}\equiv\rho_{\rm B}/\rho_c$ and radiation $\Omega_{\rm r}a^{-4}\equiv\rho_{\rm R}/\rho_c$, the model differs from the standard $\Lambda$CDM paradigm
\begin{equation}
\label{Hz}
H(a)=H_0\sqrt{\Omega_{\rm m}a^{-3} + \Omega_{\rm r}a^{-4} + \Omega_\Lambda}\,,
\end{equation}

with $\Omega_{\rm m}\equiv\Omega_{\rm b}+\Omega_{\rm DF}-\Omega_\Lambda$ and $\Omega_\Lambda\equiv-P/\rho_c$ being matter and cosmological constant $\Lambda$ parameters, respectively. At $a\equiv 1$ it holds that the sum of the density parameters is $\Omega_{\rm m} + \Omega_{\rm r} + \Omega_\Lambda\equiv1$.

\item[-] {\bf The logotropic approximation:}\\ If $K_0$ weakly varies with $P$, so $K_{0}^{\prime}\approx0$, we obtain a pure logotropic model $P\approx K_0\ln(V/V_0)$ \cite{2015EPJP..130..130C,2021PhRvD.104b3520B} or, equivalently,
\begin{equation}
\label{appr3}
P \approx -A_\star \ln \left(\frac{\rho_\star}{\rho}\right)\,.
\end{equation}
After simple algebra, we obtain \cite{2015EPJP..130..130C},
\begin{subequations}
\label{apprGL}
\begin{align}
\label{eq2}
\rho(a) &= \rho_{\rm m}a^{-3} + \rho_\Lambda\left(1 + 3B\ln a\right)\,,\\
B &\equiv [\ln(\rho_\star/\rho_{\rm m})-1]^{-1}\,,\\
A_\star&\equiv B\rho_\Lambda\,.
\end{align}
\end{subequations}
Dividing by $\rho_c$ and including the radiation, we have
\begin{equation}
\label{Hz_AS_logo}
H(a)=H_0\sqrt{\Omega_{\rm m} a^{-3} + \Omega_{\rm R} a^{-4} +  \Omega_\Lambda\left(1+3B\ln a \right)},
\end{equation}
where the flat prior ensures the condition $\Omega_{\rm m} + \Omega_{\rm r} + \Omega_\Lambda\equiv1$.

Depending on how we defined the reference density, $\rho_\star$, we can have two submodels, hereafter named GL1 and GL2, namely,
\begin{itemize}
\item[-] GL1, with $\rho_\star$ assumed to be the Planck density, $\rho_{\rm P}=5.155\times10^{96}$~kg/m$^3$ \cite{2016PhLB..758...59C} or
\item GL2, if $\rho_\star$ is a free model parameter.
\end{itemize}

\item[-] {\bf The GCG approximation:}\\ Assuming $K_0\ll K_0^\prime P$, with the substitutions $K_0^{\prime}\rightarrow -\alpha$ and $K_0\rightarrow A_\star$, we arrive at the GCG EoS \cite{2002PhRvD..66d3507B}
\begin{equation}\label{Pch}
P=-\frac{A_\star}{\alpha} \left(\frac{\rho_\star}{\rho}\right)^\alpha\,.
\end{equation}
Thus, the continuity equation gives \cite{2022EPJC...82..582Z}:
\begin{subequations}
\label{apprGCG}
\begin{align}
\Omega_{\rm DF}(a)&= \Omega_{\rm DF} \left[A_s +(1-A_s)a^{-3(1+\alpha)}\right]^{\frac{1}{1+\alpha}}\,,\\
A_s&\equiv 1 - \left(\frac{\Omega_{\rm m}-\Omega_{\rm b}}{1-\Omega_{\rm b}}\right)^{1+\alpha}= \frac{A_\star}{\alpha \rho_\star}\left[\frac{\rho_\star}{\rho(a_0)}\right]^{1+\alpha}\,.
\end{align}
\end{subequations}
Including baryonic matter and radiation, we get,
\begin{equation}
\label{Hzch}
H(a)=H_0\sqrt{\Omega_{\rm b}a^{-3} + \Omega_{\rm r}a^{-4} + \Omega_{\rm DF}(a)}\,,
\end{equation}
with the flat Universe prior $\Omega_{\rm b} + \Omega_{\rm r} + \Omega_{\rm DF}\equiv1$.
\end{itemize}
All the above scenarios describe different epochs in our Universe's evolution. To explore each of them, we constrain our models to fix the bounds over the free parameters using numerical methods as described below, in Section \ref{sezione3}.

\section{Numerical constraints}\label{sezione3}
We here constrain the model parameters via MCMC analyses. We use low and intermediate redshift catalogs: OHD \citep{2022LRR....25....6M}, SNe Ia \cite{2018ApJ...859..101S}, growth factor and RMS mass fluctuations \cite{Paul} to compute the total-log likelihood function, given by
\begin{equation}
 \ln{\mathcal{L}} = \ln{\mathcal{L}_{\rm O}} + \ln{\mathcal{L}_{\rm S}} + \ln{\mathcal{L}_{\rm f}} + \ln{\mathcal{L}_\sigma}\,,
\end{equation}
whose contributions are defined below.
\begin{itemize}
\item[-] {\bf Hubble rate data.}
Spectroscopic measurements of passively evolving galaxies provide their age $\Delta t$ and redshift $\Delta z$ differences and, consequently, cosmology-independent estimates of the Hubble rate via $H(z)\approx-(1+z)^{-1}\Delta z/\Delta t$, \cite{2002ApJ...573...37J}.
The corresponding log-likelihood function is then given by,
\begin{equation}
\label{chisquared1}
 \ln{\mathcal{L}_{\rm O}} = -\frac{1}{2} \sum_{i=1}^{N_{\rm O}} \left\{\ln{(2\pi\sigma_{H_i}^2)} + \left[\frac{H_i-H(z_i)}{\sigma_{H_i}}\right]^2\right\},
\end{equation}
where $N_{\rm O}=32$ are the OHD data points \cite{2022LRR....25....6M}.
\item[-] {\bf SNe Ia.}
The \emph{Pantheon} data set with $1048$ SNe Ia \cite{2018ApJ...859..101S} is equivalently described (in a spatially-flat Universe) by a catalog of $N_{\rm S}=6$ measurements of $E(z)\equiv H(z)/H_0$ \cite{2018ApJ...853..126R}.
The corresponding log-likelihood function is
\begin{equation}
\label{loglikeSN}
\ln \mathcal{L}_{\rm S} = -\frac{1}{2}\sum_{i=1}^{N_{\rm S}} \left\{ \mathcal E(z_i)^{\rm T} \mathbf{C}_{\rm S}^{-1}
\mathcal E(z_i) + \ln \left(2 \pi |{\rm C}_{\rm S}| \right) \right\}\,,
\end{equation}
where we defined $\mathcal E(z_i)\equiv E_i^{-1} - E(z_i)^{-1}$, the covariance matrix $\mathbf{C}_{\rm S}$ and its determinant $|{\rm C}_{\rm S}|$.
\item[-] {\bf Matter growth factor.}
Structure formation requires small initial matter density perturbations to occur.
In the linear regime, assuming homogeneity and isotropy \cite{weinberg1972gravitation}, the growth of such perturbations is described by the evolution of the density contrast $\delta=\delta\rho_{\rm m}/\rho_{\rm m}$ with respect to $\ln a$ \cite{2021PhRvD.104b3520B}
\begin{equation}\label{eq:gf}
 \delta^{\prime\prime} + \left(S\delta\right)^{\prime} +
 \left(2+\frac{H^{\prime}}{H}-\frac{T^{\prime}}{T}\right)
 (\delta^{\prime}+S\delta) - \frac{TW}{2}\delta = 0\,,
\end{equation}
where, from now on, with the prime symbol we define the derivative $\delta^\prime=d\delta/d\ln{a}$ and introduce the functions $S=3(s-w)$, $T=1+w$, $W=3(1+3s)\Omega(a)$. The adiabatic sound speed $c_{\rm s}^2=s$, the dimensionless effective matter component density of the perturbed fluid $\Omega(a)$ and the barotropic index $w$, respectively, are given by
\begin{subequations}
\label{auxiliaryfunc}
\begin{align}
    s&=\left(\frac{\partial P}{\partial a} \right) \left(\frac{\partial\rho_{\rm m}}{\partial a}\right)^{-1}\,,\\
    \Omega(a) &= \frac{\rho_{\rm m}(a)}{E(a)^2}\,,\\
    \label{ad_sound_speed}
    w(a) & = -1-\frac{2}{3}\frac{H^\prime(a)}{H(a)}\,,
\end{align}
\end{subequations}
where in $E(a)$ the contribution of the radiation is neglected.

Using the growth factor $f=(\ln{\delta})^\prime$, Eq.~\eqref{eq:gf} reads
\begin{equation}
\label{eqn:exact_f}
f^{\prime}+ \left(2 + f + \frac{H^{\prime}}{H}-\frac{T^{\prime}}{T}\right)\left(f+S\right) + S^\prime - \frac{TW}{2}= 0\,.
\end{equation}
So, Eq. \eqref{eqn:exact_f} can be solved by assuming a phenomenological expression of the growth factor in the form of $f \approx \Omega(a)^\gamma$ \cite{Paul}, where $\gamma$ is the growth index.
The log-likelihood from $N_{\rm f}=11$ data is given by
\be\label{flikelihood}
 \ln{\mathcal{L}_{\rm f}} = -\frac{1}{2}\sum_{i=1}^{N_{\rm f}}\left\{\ln{\left(2\pi\sigma_{\rm f_i}^2\right)} + \left[\frac{f_{\rm i}-f\left(z_i\right)}{\sigma_{\rm f_i}}\right]^2\right\}\,.
\ee
\item[-] {\bf RMS mass fluctuations.}
An alternative probe of $\delta(z)$ is the RMS mass fluctuation $\sigma_8(z)$:
\begin{equation}
 \sigma_8(z) = \sigma_8(0)\frac{\delta(z)}{\delta(0)} = \sigma_8(0) \exp\left[\int^{\frac{1}{1+z}}_{1}\Omega(a)^\gamma\mathrm{d}a\right]\,,
\end{equation}
where $\sigma_8(0)$ is the value at $z=0$. Most of the available data originate from the redshift evolution of the flux power spectrum of the Ly--$\alpha$ forest \cite{Paul}. To avoid the fitting of $\sigma_8(0)$, we use
\begin{equation}
 s_8(z^i_{i+1}) = \frac{\exp\left[\int^{\frac{1}{1+z_i}}_{1}\Omega(a)^\gamma\mathrm{d}a\right]}{\exp\left[\int^{\frac{1}{1+z_{i+1}}}_{1}\Omega(a)^\gamma\mathrm{d}a\right]}\,.
\end{equation}
The corresponding log-likelihood is
\bea
 \ln\mathcal{L}_\sigma = -\frac{1}{2}\sum_{i=1}^{N_\sigma}\left\{\ln (2\pi\sigma_{8,i}^2) +\left[\frac{s_{\rm 8,i}-s_8(z^i_{i+1})}{\sigma_{8,i}}\right]^2\right\}\label{chisquared2}
\eea
where $N_\sigma=17$ and $\sigma_{\rm s_{8,i}}$ is derived by error propagation from the errors of $s_8(z_i)$ and $s_8(z_i+1)$.
\end{itemize}
\begin{table*}
\centering
\setlength{\tabcolsep}{.5em}
\renewcommand{\arraystretch}{1.1}
\caption{Best-fit and derived parameters (with $1$--$\sigma$ errors), and statistical tests of $\Lambda$CDM, GL1, GL2, and GCG models.}
\begin{tabular}{lccccccccrcc}
\hline\hline
                                    &
\multicolumn{4}{c}{Best-fit parameters}&
                                    &
\multicolumn{2}{c}{Derived parameters}&
                                    &
\multicolumn{3}{c}{Statistical tests}      \\
\cline{2-5}\cline{7-8}\cline{10-12}
Model                               &
$h_0$                               &
$\Omega_{\rm m}$                    &
$B$                                 &
$\alpha$                            &
                                    &
$\log(\rho_\star/\rho_c)$           &
$\log(A_\star/\rho_c)$              &
                                    &
$-\ln \mathcal L$                   &
AIC (BIC)                           &
$\Delta$                           \\
\hline
$\Lambda$CDM                        &
$0.694^{+0.028}_{-0.026}$           &
$0.293^{+0.030}_{-0.033}$           &
--                                  &
--                                  &
                                    &
$122.76$                            &
$-2.602^{+0.030}_{-0.032}$          &
                                    &
$97.80$                             &
$200$ ($204$) & $0$ ($0$)           \\
GL1                                &
$0.692^{+0.029}_{-0.025}$           &
$0.294^{+0.028}_{-0.034}$           &
--                                  &
--                                  &
                                    &
$122.76$                            &
$-2.603^{+0.031}_{-0.033}$          &
                                    &
$97.74$                             &
$200$ ($204$) & $0$ ($0$)           \\
GL2                                &
$0.693^{+0.025}_{-0.032}$           &
$0.322^{+0.039}_{-0.042}$           &
$0.040^{+0.078}_{-0.039}$           &
--                                  &
                                    &
$<12.18$                            &
$<-2.40$                            &
                                    &
$97.47$                             &
$201$ ($208$) & $1$ ($4$)         \\
GCG                                 &
$0.690^{+0.028}_{-0.027}$           &
$0.316^{+0.054}_{-0.044}$           &
--                                  &
$0.10^{+0.18}_{-0.15}$              &
                                    &
$122.76$                            &
$-13.275^{+0.037}_{-0.041}$         &
                                    &
$97.25$                             &
$201$ ($207$) & $1$ ($3$)           \\
\hline
\end{tabular}
\label{tab:results}
\end{table*}
\begin{figure*}
{\includegraphics[width=0.27\hsize,clip]{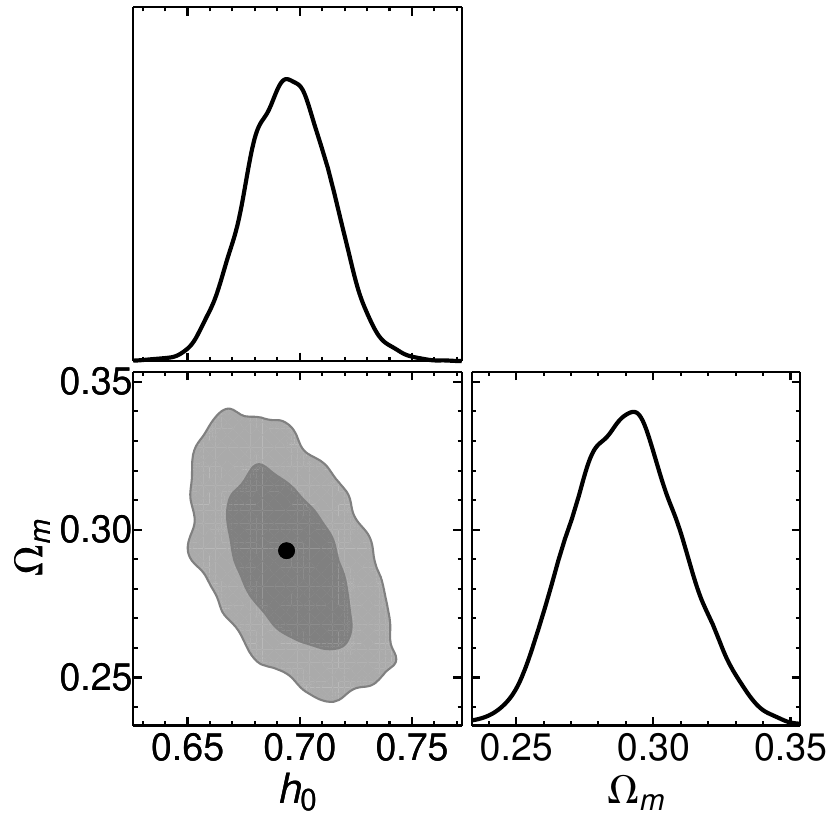}\hfill
\includegraphics[width=0.27\hsize,clip]{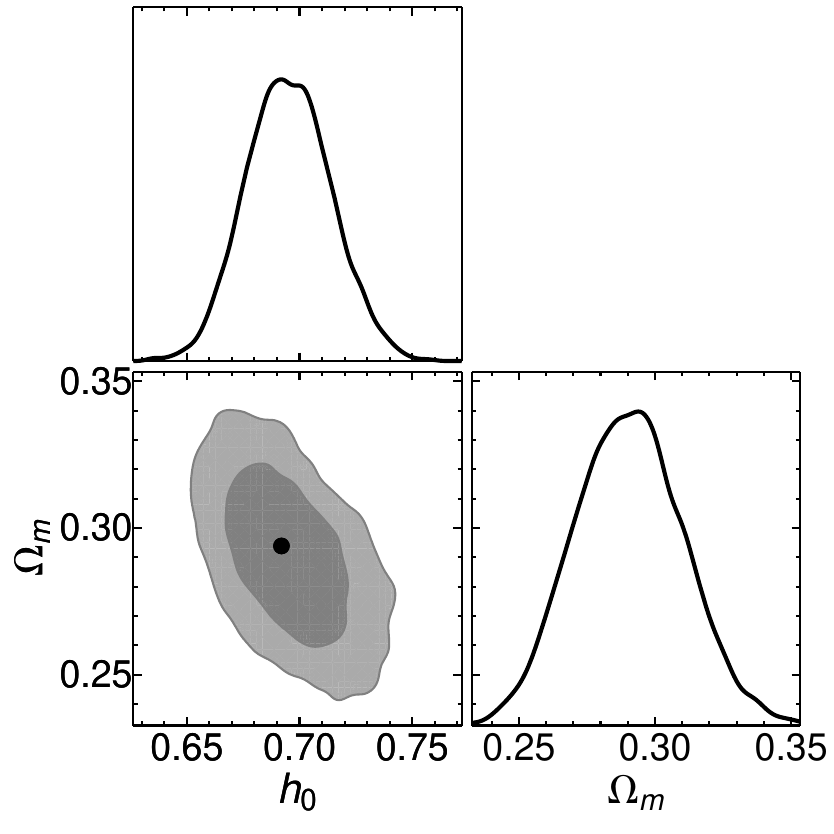}\hfill}

{\includegraphics[width=0.38\hsize,clip]{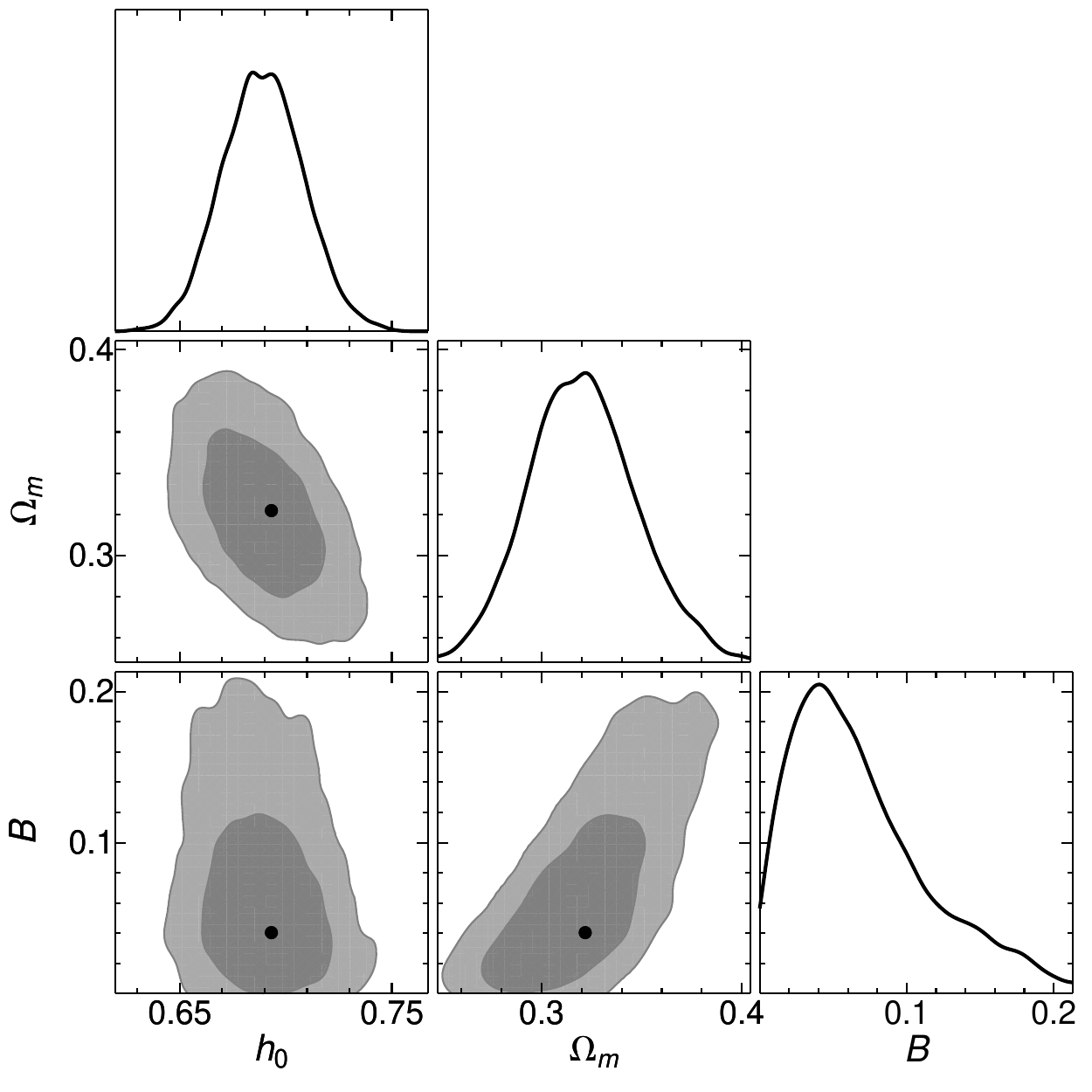}\hfill
\includegraphics[width=0.38\hsize,clip]{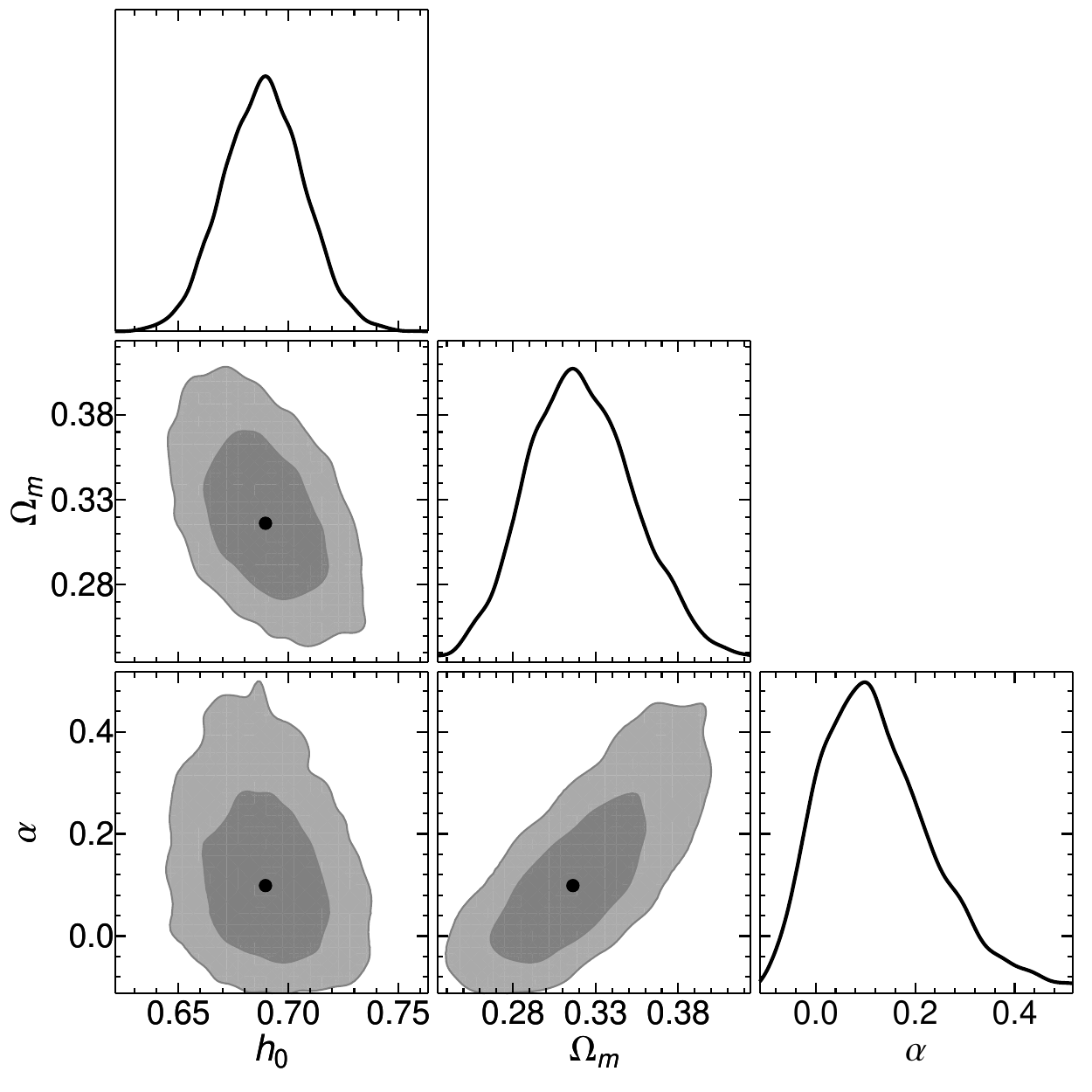}\hfill}
\caption{Contour plots and best-fit parameters (black dots) of $\Lambda$CDM (top left), GL1 (top right), GL2 (bottom left), and GCG (bottom right) models. Darker (lighter) areas mark $1$--$\sigma$ ($2$--$\sigma$) confidence regions.}
\label{fit:contours}
\end{figure*}
Fig.~\ref{fit:contours} and Tab. \ref{tab:results} show, respectively, the contour plots and the best-fit parameters of the considered submodels. We indicated $H_0=100h_0$, and fixed $\Omega_{\rm r}=9.265\times10^{-5}$ and  $\Omega_{\rm b}=0.0493$ \cite{Planck2018}.

With $N$ data, maximum log-likelihood $\ln \mathcal L$ and parameters $p$, for each proposed submodel we define the \emph{Aikake Information Criterion} (AIC) and the \emph{Bayesian Information Criterion} (BIC) \cite{2007MNRAS.377L..74L}, respectively,
\begin{subequations}
\label{stattests}
\begin{align}
{\rm AIC}&=-2\ln \mathcal L+2p\,,\\
{\rm BIC}&=-2\ln \mathcal L+p\ln N\,.
\end{align}
\end{subequations}
The best-suited model provides the lowest values AIC$_0$ and BIC$_0$.
The differences $\Delta={\rm AIC}-{\rm AIC}_0$ (${\rm BIC}-{\rm BIC}_0$) provide evidence against the proposed model as follows
\begin{itemize}
    \item[-] $\Delta\in[0,\,3]$, weak evidence;
    \item[-] $\Delta\in (3,\,6]$, mild evidence;
    \item[-] $\Delta>6$, strong evidence.
\end{itemize}
Immediately, Tab. \ref{tab:results} shows that $\Lambda$CDM and GL1 models are equally best suited, the GCG model is weakly disfavoured, and the GL2 model is mildly excluded. These results also enable constraints on the parameters ($\alpha$, $\rho_\star$, $A_\star$).
\begin{itemize}

    \item[-] {\bf $\Lambda$CDM}. Assuming $\alpha\gtrsim0$, from Eqs. \eqref{apprLCDM} we get
    \begin{equation}
         A_\star = \frac{3H_0^2}{8\pi G}\left[\ln\left(\frac{8\pi G\rho_\star}{3H_0^2}\right)\right]^{-1}\left(1-\Omega_{\rm M} - \Omega_{\rm R}\right)\,,
    \end{equation}
    that is not enough to break the degeneracy between  $\rho_\star$ and $A_\star$, unless we impose $\rho_\star\equiv\rho_{\rm P}$ as in Tab.~\ref{tab:results}.
    \item[-] {\bf GL1} and {\bf GL2}. Assuming $\alpha\gtrsim0$, from Eqs.~\eqref{apprGL} we are now able to set conditions on both $\rho_\star$ and $A_\star$
    \begin{subequations}
    \begin{align}
    \label{rhostar}
    \rho_\star=&\frac{3H_0^2\Omega_{\rm M}}{8\pi G}\exp\left(1+\frac{1}{B}\right)\,,\\
        A_\star=&\frac{3H_0^2B}{8\pi G}\left(1-\Omega_{\rm m} - \Omega_{\rm r}\right)\,.
    \end{align}
    \end{subequations}
    For GL1 we evaluated only $A_\star$, since $\rho_\star\equiv\rho_{\rm P}$, while for GL2 we evaluated both $\rho_\star$ and $A_\star$ (see Tab.~\ref{tab:results}).
    \item[-] {\bf GCG}. Here $\alpha$ is a model parameter and its best-fit value, see Tab.~\ref{tab:results},  agrees with Ref.~\cite{2022EPJC...82..582Z}.
    However, from Eqs.~\eqref{apprGCG} we obtain  one condition only
        \begin{equation}
        \label{Astar_GCG}
         A_\star = \alpha\frac{A_{\rm s}}{\rho_\star^\alpha}\left(\frac{3H_0^2}{8\pi G}\right)^{1+\alpha}\left(1-\Omega_{\rm b} - \Omega_{\rm r}\right)^{1+\alpha}\,,
    \end{equation}
that does not allow us to ``disentangle" $\rho_\star$ and $A_\star$.
\end{itemize}

\section{Physical interpretation}\label{sezione4}
Some interesting physical consequences can be drawn from the above analyses.
\begin{itemize}
    \item[-] At $a\approx1$, the Murnaghan EoS in Eq. \eqref{P1} is approximated by the $\Lambda$CDM model for $\rho\approx\rho_{\rm c}$.
    \item[-] GL1 is statistically identical to and degenerates with the $\Lambda$CDM model; GL2 is mildly excluded. As $\Lambda$CDM and GL1 models are limiting cases of Eq. \eqref{P1}, we deduce that the condition $\rho_\star\equiv\rho_{\rm P}$ holds also for the most general Murnaghan EoS, i.e., when no approximation holds.
    \item[-] At $10^{-5}\lesssim a\lesssim 1$, for $\rho_\star\equiv\rho_{\rm P}$, the term $(\rho_{\rm P}/\rho)^\alpha$ dominates and Eq.~\eqref{P1} is approximated by the GCG model, which is the only framework able to get constraints on $\alpha$ (see Tab.~\ref{tab:results}).
    \item[-] Using the constraints on $\rho_\star\equiv\rho_{\rm P}$ (from $\Lambda$CDM and GL1 submodels) and $\alpha$ (from the GCG model), from Eq. \eqref{Astar_GCG} we get the value of $A_\star$ for the GCG case (see Tab.~\ref{tab:results}) that also holds for the total Murnaghan EoS.
    \item[-] At early times, when $a\approx0$ and $x=\rho_{\rm P}/\rho\approx1$, at the lowest order we have $x^\alpha - 1\approx\ln x^\alpha$, thus, Eq.~\eqref{P1} reduces to the GL1 model, as in  Eq.~\eqref{appr3}.
\end{itemize}
\begin{figure}
\includegraphics[width=\hsize,clip]{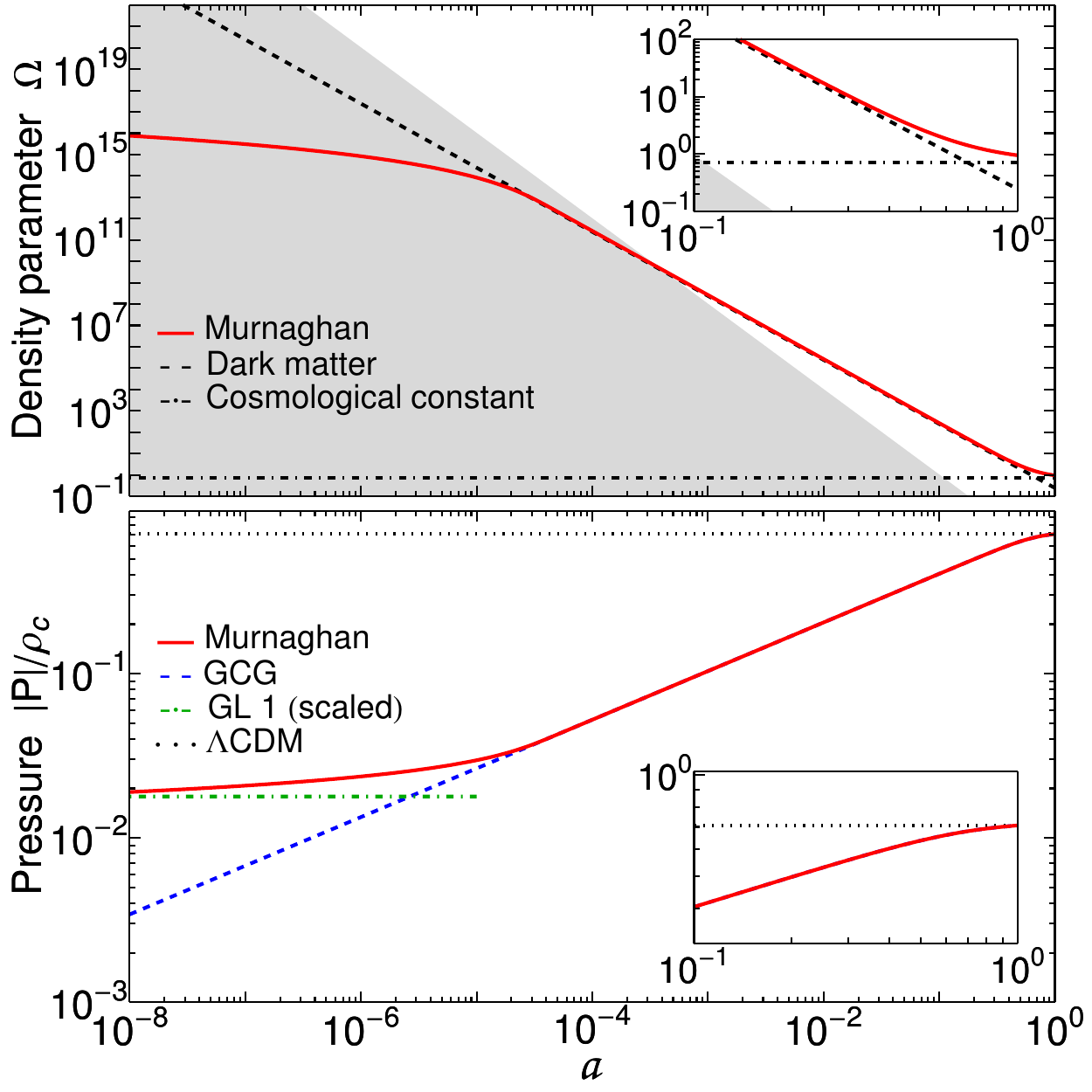}
\caption{\emph{Top panel}: Murnaghan density compared to $\Lambda$CDM dark matter, dark energy, and radiation (shaded area) ones. \emph{Bottom panel}: comparison among Murnaghan, GCG, GL1 (rescaled to Murnaghan as $a\rightarrow0$), and $\Lambda$CDM EoSs. The insets focus on $0.1\leq a\leq1$. Parameters are taken from Tab.~\ref{tab:results}.}
\label{fig:EoS}
\end{figure}
In view of the above considerations, we fix the following viable bounds:
\begin{subequations}
\label{consMurnpar}
\begin{align}
    \bar \alpha&=0.10^{+0.18}_{-0.15}\,,\\
    \bar\rho_\star&=\rho_{\rm P}\,,\\
    \bar A_\star&= \left(5.31^{+0.48}_{-0.48}\right)\times10^{-14}\rho_{\rm c}\,,
    \end{align}
\end{subequations}
which can be considered as ``average quantities" obtained from the limiting cases at different times.

The corresponding best-fit cosmological parameters, with associated errors, can be found from the numerical bounds obtained from $\Lambda$CDM, GL1 and GCG models:
\begin{subequations}
\label{consMurncos}
\begin{align}
    \bar h_0&=0.692^{+0.030}_{-0.029}\,,\\
    \bar\Omega_{\rm M}&=0.301^{+0.069}_{-0.041}\,.
\end{align}
\end{subequations}
The above model ($\bar\alpha$, $\bar\rho_\star$, $\bar A_\star$) and cosmological ($\bar h_0$, $\bar\Omega_{\rm m}$) parameters have been used to obtain the numerical density displayed in the top panel of Fig.~\ref{fig:EoS}.
The comparison among Murnaghan, GCG, GL1 (rescaled to Murnaghan as $a\rightarrow0$) and $\Lambda$CDM scenarios is portrayed in the bottom panel of Fig.~\ref{fig:EoS}.  These curves show explicitly the behavior of the full numerical solution and enable us to summarize our physical interpretation as follows.
\begin{itemize}
\item[-] At early times, or $a\lesssim10^{-5}$, the model acts as a logotropic fluid that degenerates with the $\Lambda$CDM model. Thus, in our framework we expect that structure formation proceeds with a behaviour similar to the current standard cosmological model.
\item[-] At $10^{-5}\lesssim a\lesssim 1$, the model acts as a GCG solution and agrees with late time observations\footnote{Possible departures can be detected adopting high-distance indicators, such as gamma-ray bursts.}.
\item[-] In the current epoch, or at $a\approx1$, the model behaves again as GL1 or $\Lambda$CDM models.
\end{itemize}

Thus, the overall model appears to extend both logotropic and GCG scenarios, quite agreeing with observations, explaining in which regions of the Universe evolution the $\Lambda$CDM model is recovered and, consequently, fixing the coincidence issue between $\Omega_\Lambda$ and $\Omega_{{\rm m}}$.

To disclose the impact on structure formation, we analyze below how our model acts over linear perturbations.

\section{Impact on linear perturbations}\label{sezione6}

The basic expressions used to describe structure formation in the Universe are the continuity and Euler equations, respectively given by
\begin{subequations}
\begin{align}
\label{eq:continuity}
\delta^{\prime}+3(s-w) \delta+(1+w) \tilde{\theta}&=0,\\
\label{eq:euler}
\tilde{\theta}^\prime + \left(2+\frac{H^\prime}{H}\right)\tilde{\theta}+\frac{3}{2}(1+3s)\Omega(a)\delta&=0,
\end{align}
\end{subequations}
where we defined $\tilde{\theta}=\theta/H$ and introduced the divergence $\theta$ of the peculiar velocity $\mathbf{u}$, such that $\theta=\vec{\nabla}\cdot \mathbf{u}$.
All the other quantities involved in the above equations are already defined by Eqs. \eqref{auxiliaryfunc} in Section \ref{sezione3}.
\begin{figure}
\includegraphics[width=0.8\hsize,clip]{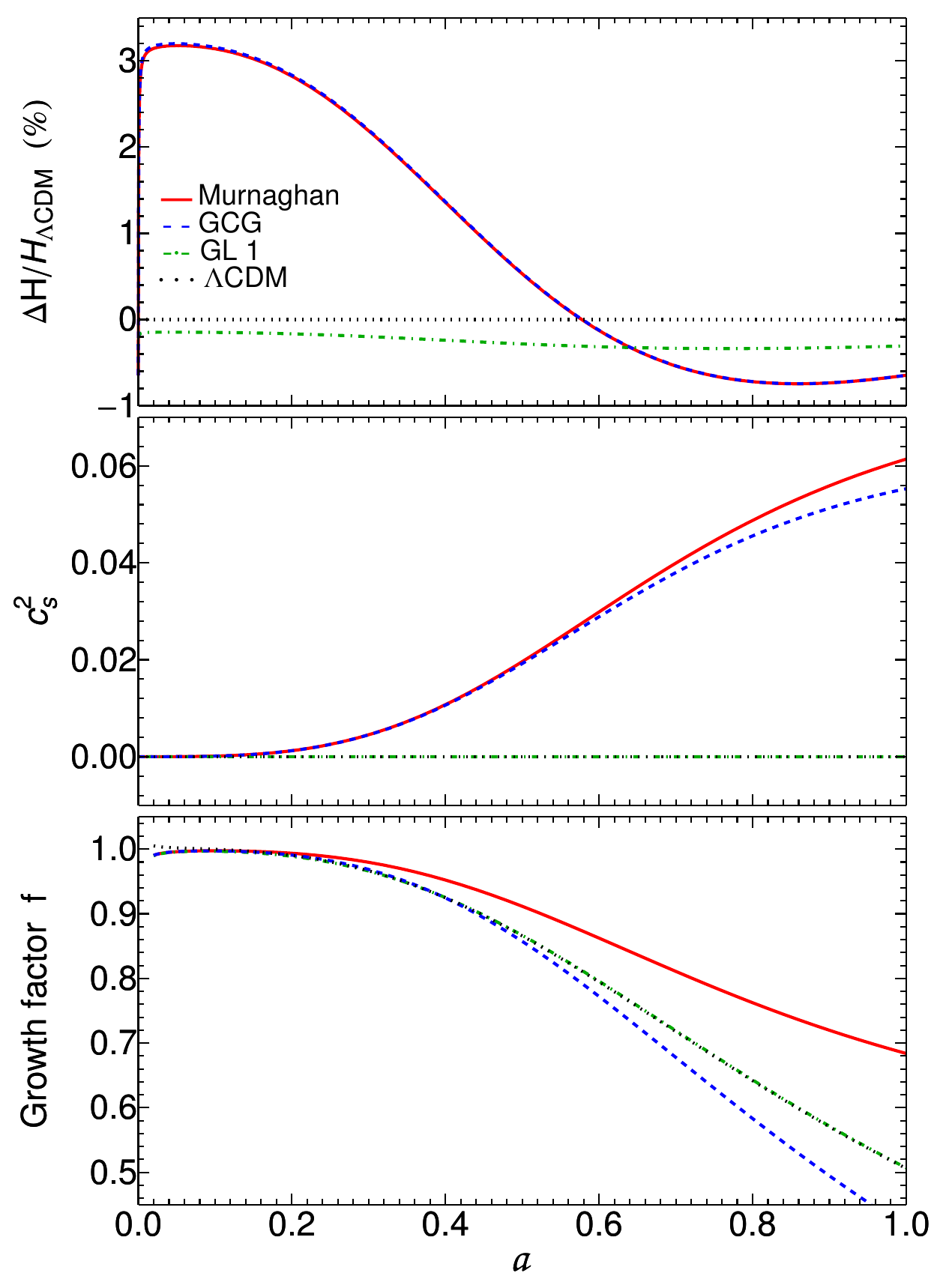}
\caption{Murnaghan, GCG, GL1, and $\Lambda$CDM models comparison. \emph{Top panel}: Hubble rate relative differences with respect to the $\Lambda$CDM one. \emph{Middle panel}: squared sound speeds. \emph{Bottom panel}: growth  factors. Parameters are taken from Tab.~\ref{tab:results}.}
\label{fig:pert}
\end{figure}
By taking the derivative of Eq.~\eqref{eq:continuity} and substituting it in Eq.~\eqref{eq:euler} we derive Eq. \eqref{eq:gf} and then Eq. \eqref{eqn:exact_f}, that can be solved following the standard assumptions listed below \cite{2021PhRvD.104b3520B}.
\begin{itemize}
    \item[1.] The dark matter is the clustering component, so one can set $w=0$.
    \item[2.] With the phenomenological solution $f \approx \Omega_{\rm m}^{\gamma}(a)$ \citep{Paul}, we obtain a first-order differential equation for $\gamma$ which is linearized by assuming $\Omega_{\rm m}\approx\mathcal{O}(1)$.
    \item[3.] The sound speed vanishes, so one can set $s=0$.
\end{itemize}
However, in our model we have the advantage to just consider the first assumption above and proceed with the numerical evaluations for $\Lambda$CDM, GL1, GCG approximations and Murnaghan model, we obtain the evolution of the Hubble parameter in Fig. \ref{fig:pert} (top panel),  portrayed as the relative departure with respect to the reference $\Lambda$CDM model that never exceeds the $\approx3\%$ error for $0\leq a\leq1$. The evolution of the adiabatic sound speed, which determines the stability and the validity of a given model against perturbations, is shown in Fig. \ref{fig:pert} (middle panel).
The evolution of the growth factor $f$ is shown in the bottom panel of Fig.~\ref{fig:pert}.
The parameters used in the plots are taken from Tab.~\ref{tab:results} whereas the parameters for the Murnaghan model are taken from Eqs.~\eqref{consMurnpar}--\eqref{consMurncos}.

Our main results are thus listed in the following.
\begin{itemize}
\item[-] All the models have negligible, positive and real sound speeds (identically zero only for the $\Lambda$CDM model).
\item[-]  At very early times, for all models, $f$ flattens to a value $\approx1$, as expected, then decreases.
\item[-]  At late times, the GCG (GL1) model departs from (degenerates with) the $\Lambda$CDM one due to its non-zero, albeit extremely small,  sound speed.
\item[-]  Our model largely deviates from the $\Lambda$CDM scenario, resulting in a \emph{more efficient} growth of perturbations.
This cannot be explained by the sound speed, similar to that of the GCG model (see the middle panel of Fig.~\ref{fig:pert}), but rather depends on the density parameter of the perturbed fluid $\Omega(a)$, that accounts for the total dark fluid density enhancing the growth of perturbations.
\end{itemize}

Hence, our model, that employs the novel concept of matter with non-zero pressure, seems to better behave than the standard $\Lambda$ model alone, showing how dark energy can change its functional behaviours throughout the entire Universe evolution.

\section{Conclusions}\label{sezione7}

We are currently witnessing increased interest in \emph{unified dark energy} models, aiming to explain cosmic acceleration using a single matter fluid exhibiting varying properties at different stages of the Universe's evolution. Within this context, we propose employing a double polytropic EoS, consisting of two adiabatic indexes, one set to zero, while the other is negative. This approach allows us to replicate the effects of a specific EoS, resembling a combination of a Chaplygin gas fluid and a non-zero cosmological constant contribution. However, this combination seems finely tuned and compatible with the bare cosmological constant value, but not with quantum fluctuations. In this regard, we showed the fluid exhibits similarities to the Murnaghan EoS, a widely-used relation in solid state physics, particularly in describing pressure regimes within crystals.

The Murnaghan EoS is therefore here introduced in cosmological contexts. In particular, we showed that it reduces to the pure $\Lambda$CDM, logotropic and GCG cases, depending on the choices of $\rho$. This interesting property is a crucial feature of our model, showing that it represents a \emph{unified dark energy framework} that at the same time, \emph{unifies the phenomenological descriptions of different dark energy models into one single fluid of matter with non-zero pressure, found in physical contexts and applied to the observable Universe}.

We considered characteristics during the background cosmological regime and at early times, conducting a thorough investigation that encompassed cosmological observations involving SNe Ia, OHD, matter growth factor and $\sigma_8$ data. Our emphasis is centered on structure formation, with a specific focus on first-order perturbation theory.

We found that our model presents an intriguing prospect, with improved behavior when compared to the $\Lambda$CDM model, which uniformly assumes a vanishing sound speed for perturbations. In contrast, our model introduces a non-zero cutoff contingent upon the density scale. This particular attribute prompts an exploration into whether galaxies could potentially form even earlier than standard expectations, partially in line with recent developments recently discussed by the James Webb Telescope \cite{2023Natur.622..707A}.

Future work will concentrate on more intricate applications of the Murnaghan EoS, incorporating self-interacting fluids. As demonstrated in Ref. \cite{Dunsby:2023qpb}, it is plausible to interpret this fluid through a scalar field description, leading to a symmetry-breaking potential that contributes to the constant term evident in the Murnaghan EoS. Consequently, the incorporation of more complex versions of the potential itself will offer insights into potential refinements of our current model, positioning it as a robust alternative to prior unified dark energy models.

\section*{Acknowledgements}
The work of OL is  partially financed by the Ministry of Education and Science of the Republic of Kazakhstan, Grant: IRN AP19680128. MM acknowledges the support of INFN, iniziativa specifica MoonLIGHT for financial support. The work by MM is partially financed by the Ministry of Education and Science of the Republic of Kazakhstan, Grant: IRN BR21881941. PKSD acknowledges the First Rand Bank (SA) for financial support.

\end{document}